\documentstyle[prd,preprint,aps,epsfig,floats,axodraw]{revtex}
\begin{document}

\tightenlines
\input epsf.tex
\def\DESepsf(#1 width #2){\epsfxsize=#2 \epsfbox{#1}}
\draft
\thispagestyle{empty}
\preprint{\vbox{ \hbox{UMDPP-02-002}
\hbox{July 2001}}}
\title{\Large \bf Connecting Bimaximal Neutrino Mixing to a Light
Sterile Neutrino }

\author{\large\bf
R.N. Mohapatra\footnote{e-mail:rmohapat@physics.umd.edu}}

\address{ Department of Physics, University of Maryland\\
College Park, MD 20742, USA}

\maketitle

\thispagestyle{empty}

\begin{abstract}

It is shown that if small neutrino masses owe their origin to the
conventional seesaw mechanism and the MNS mixing matrix is in the exact
bimaximal form, then there exist symmetries in the theory that allow
one of the righthanded neutrinos to become naturally massless, making it
a candidate for the sterile neutrino discussed in the literature.
Departures from the exact bimaximal limit leads to tiny mass for the
sterile neutrino as well as its mixing to the active neutrinos. This
provides a minimal theoretical framework where a simultaneous explanation
of the solar, atmospheric and LSND observations within the so-called 3+1
scenario may be possible.
\\[1ex] PACS: {14.60.Pq;
14.60.St; 11,10.Kk;}
 \end{abstract}

\newpage

\section{Introduction}
At the moment, there are three classes of experiments that provide
positive indications for neutrino oscillations: solar neutrino data 
from seven different experiments, Chlorine, Kamiokande,
Super-Kamiokande, SAGE, GALLEX, GNO, SNO, all experiments giving evidence
for missing electron neutrinos emitted from the solar core,
which can be understood in terms of neutrino
oscillations\cite{ref1}; atmospheric neutrino data from Super-Kamiokande
which has not only confirmed earlier indications of a muon neutrino
deficit but also has provided more precise data that has established at a
high confidence level that the muon neutrinos from the cosmic rays are
oscillating to tau neutrinos\cite{ref2}. These two evidences are on very
solid footing. The third piece of evidence is from the Los Alamos LSND
experiment that
shows an oscillation from the muon neutrino to the electron
neutrino. This experiment however has neither been been confirmed nor
refuted by other experiments\cite{lsnd}. The
KARMEN\cite{karmen} experiment which looked for $\nu_{\mu}-\nu_e$
oscillation did not find any evidence for it and eliminated a large
fraction of the parameter space allowed by LSND. It is hoped
that the Mini-BOONE experiment at FERMILAB will settle the issue in near
future.

 Theoretical analyses have made it clear that explanation of the
observations in terms of neutrino oscillations require that the three
mass differences $\Delta m^2_{\nu_e-x}$ for the solar case, $\Delta
m^2_{\nu_{\mu}-\nu_{\tau}}$ for the atmospheric case and $\Delta
m^2_{\nu_e-\nu_{\mu}}$ for explaining LSND are all of very different
orders of magnitude. This poses a theoretical problem for models with only
three active neutrinos since with three neutrinos, one can at most have
 two different mass differences. The simplest way to provide
a simultaneous understanding of all the above mentioned
oscillation data appears to be to postulate the existence of an additional
ultralight neutrino, which in order to be consistent with the LEP data
must be a sterile neutrino\cite{cald}.

In the presence of a sterile neutrino, there are several ways to
understand the observations. We mention only two of them, one called 2+2
scheme\cite{cald} and another called 3+1 scheme\cite{3p1} in the
literature. The 2+2 scheme has the $\nu_{\mu,\tau}$ neutrinos with mass
around an eV and $\nu_{e,s}$ with mass near $10^{-3}$ eV, with the later
explaining the solar neutrino data, the former explaining the
atmospheric neutrino data and the gap between the two pairs explaining
the LSND results. Recent SNO data disfavors the original version
of the model where all the missing solar $\nu_e$'s are converted via
a small
angle MSW mechanism only to the sterile neutrinos. The situation where
only a
fraction of the missing $\nu_e$s convert to $\nu_s$s and the rest to
active ones has been studied in several papers\cite{danny}. A particular
challenge in this model would be to fit both the
SNO Super-Kamiokande gap and the more or less flat neutrino energy
distribution observed in both the SNO and Super-Kamiokande experiments.

 In the 3+1 picture, on the other hand, it is assumed that the three
active neutrinos are bunched together at a small mass value (say around
$6\times 10^{-2}$ eV or so), with the sterile neutrino at a mass near
an eV. The atmospheric and solar neutrino data is explained by the
oscillations among active neutrinos whereas the LSND data is explained by
indirect oscillations involving the sterile neutrino\cite{pati}. The
advantage of
this picture in view of the SNO data is that the flat energy spectrum is
explained by postulating bimaximal\cite{bimax} MNS mixing pattern among
the active
neutrinos and using the large mixing angle MSW solution. The
SNO-Super-Kamiokande gap 
is understood in terms of the neutral current interactions
of the $\nu_{\mu}$'s to which the solar $\nu_e$'s convert. It must however
be pointed out that 3+1 pattern works only for certain values of the
$\Delta m^2_{LSND}$ as has been noted in ref.\cite{3p1} and there
are papers noting that it may have problem accomodating all accelerator
and reactor constraints as well as the positive signals for
oscillation\cite{cons} and yet have room in its parameter space to
explain the LSND results.

If experiments ultimately establish the existence of a sterile neutrino,
a key theoretical challenge would be to understand its origin in the
context of physics beyond the standard model, specially why its mass is so
small
even though it has no electroweak quantum numbers. Most existing scenarios
postulate new fermions beyond the conventional seesaw framework i.e. new
singlet fermions beyond the three right handed dictated by quark lepton
symmetric extension of the standard model. They are usually taken to be 
mirror neutrinos\cite{mirror}, extra singlet fermions\cite{singlet} or
modulinos\cite{smir}, $E_6$ singlets\cite{e6} or bulk
neutrinos\cite{bulk} as in models with large extra dimensions. 

Our goal in this brief note is to point out a more economical
possibility. We show that when one tries to obtain the exact bimaximal
form for the MNS matrix in the framework of the seesaw\cite{seesaw}
mechanism, there appear new symmetries of the neutrino mass matrix in
special limits that
predict that one of the right handed neutrinos, which normally would have
been superheavy, remains massless whereas the
other two are superheavy as expected in the seesaw mechanism. The massless
right handed neutrino can
therefore be identified with the sterile neutrino, providing not only a
completely new picture for the sterile neutrino but also connecting
it to the bimaximal mixing among the active neutrinos. 
  
Slight departures from the bimaximal pattern endow this
massless sterile neutrino with an ultralight mass as well as mixings with
the active neutrino, so that one can now use this to understand known
oscillation data. The resulting
picture is the 3+1 type discussed above. Thus if the 3+1 scenario stands
the test of time, the model discussed in the present paper would provide
an interesting minimal theoretical scenario that would connect two
apparently different phenomena without going beyond the usual quark lepton
symmetric framework.

\section{Symmetry reason for  bimaximal mixing sterile neutrino
connection}

Let us start by writing down the generic bimaximal MNS mixing matrix
with the definition:
\begin{eqnarray}
\nu_{\alpha}=\sum_i U_{\alpha i} \nu_i;
\end{eqnarray}
where $\nu_{\alpha}$ and $\nu_i$ are respectively the flavor and mass
eigenstates.
\begin{eqnarray}
U_{bimax}~=~\left(\begin{array}{ccc} c & -s & 0\\ \frac{s}{\sqrt{2}}
& \frac{c}{\sqrt{2}} & \frac{1}{\sqrt{2}}\\ \frac{s}{\sqrt{2}} 
& \frac{c}{\sqrt{2}} & -\frac{1}{\sqrt{2}}\end{array}\right).
\end{eqnarray}
When $c=s=\frac{1}{\sqrt{2}}$, we will refer to this as the exact
bimaximal limit alluded to above. It was shown in ref.\cite{nuss}
that the most general mass matrix involving the active neutrinos, which
leads to the strict bimaximal pattern in the basis where the charged
leptons are mass eigenstates is
\begin{eqnarray}
M_3 = \left(\begin{array}{ccc} a+b & m & m\\ m & a & b \\ m & b & a
\end{array}\right).
\end{eqnarray}
This mass matrix has a permutation symmetry $S_2$
operating on the $\nu_{\mu}$ and $\nu_{\tau}$. Let us call this 
$S_{2L}$ since it acts only on the left-handed neutrinos. In
the limit of exact $SU(2)_L$ gauge symmetry of the electroweak
interactions, this symmetry will also have consequences for the charged
lepton masses (i.e. $\mu, \tau$ masses). The difference of the $\mu$ and
the $\tau$ mass will then be attributed to its breaking. Note
that the gauge interactions are invariant under $S_{2L}$.

In the matrix $M_3$, $m, a, b$ are three free parameters. In order to
further relate the bimaximal mixing to symmetries of the leptonic world,
we will restrict ourselves to the case when $a=b=0$. In this case, the
matrix $M_3$ has the symmetry $L_e-L_{\mu} -L_{\tau}$ symmetry\cite{sym}
in addition to the $S_{2L}$ symmetry. Let us therefore consider the
combination
of $S_{2L}\times U(1)_{e-\mu-\tau}$ as a symmetry of the weak Leptonic
Lagrangian to zeroth order.

The next step in our discussion is the seesaw\cite{seesaw} mechanism which
provides an explanation of the small neutrino masses. In order to
implement this, we will include three right handed neutrinos in the theory
(to be denoted by $\nu_{eR}, \nu_{\mu R}, \nu_{\tau R}$) in the
standard model. This makes the theory quark lepton symmetric. If we denote
the
mass matrix for the right handed neutrinos as $M_R$, then the complete
$6\times 6$ mass matrix involving the Dirac mass and Majorana mass for the
neutrinos can be written as 
\begin{eqnarray}
{\cal M}_{LR}~=~ \left(\begin{array}{cc} 0 & M_D \\ M^T_D &
M_R\end{array}\right).
\end{eqnarray}
When $M_R$ is not a singular matrix, one can obtain the mass matrix for
the light neutrinos as
\begin{eqnarray}
{\cal M}_{\nu}~=~ - M^T_DM^{-1}_RM_D
\end{eqnarray}
This is the so-called type I seesaw formula(see e.g. \cite{rev}). 
On the other hand when $M_R$ matrix is singular, this means that it
has one or more zero eigenvalues and one must ``take them out'' of the
matrix before using
the seesaw formula to obtain the light neutrino mass matrix\footnote{For
earlier discussion of
singular seesaw to get light sterile neutrinos, see \cite{sing}. None of
these papers connected the existence of the sterile neutrino to the
bimamaximal mixing among neutrinos, as is done in this paper.} . In this
case, one of the three
righthanded neutrinos  will have a mass which is far below the seesaw
scale. If it has a Dirac
mass with one or more of the left handed neutrinos at the tree level, its
mass will be its Dirac mass, which, apriori, can be quite large (of order
or less than the weak scale). Without further restriction, the light right
handed neutrino in
general is not light enough to qualify as the desired ultralight sterile
neutrino needed for understanding the oscillation data. 

On the other hand, if the theory has the symmetry dictated by the mass
matrix discussed above i.e.
$U(1)_{e-\mu-\tau}\times S_{2L}\times S_{2R}$ where 
 $S_{2R}$ operates on the righthanded neutrinos, then the
Dirac mass connecting the massless righthanded neutrino to the left
handed neutrinos vanishes and the eigenvector of the righthanded
neutrino mass matrix corresponding to the zero eigenvalue emerges as a
viable candidate for the sterile neutrino. Its tree level mass is zero.

To study this in detail, note that
the symmetries of the theory force the $M_R$ to take the form
\begin{eqnarray} M_R~=~ \left(\begin{array}{ccc} 0 & M & M \\ M & 0 & 0 \\
M & 0 & 0\end{array}\right).
\end{eqnarray}
The Dirac mass for the neutrinos that connects the left and the right
handed neutrinos takes the form
\begin{eqnarray}
M_D~=~\left(\begin{array}{ccc} m_{11} & 0 & 0\\
0 & m_0 & m_0\\
0 & m_0 & m_0\end{array}\right)
\end{eqnarray}
Note that both the $M_D$ and $M_R$ have one zero eigenvalue each. They
correspond to the linear combinations
$\frac{\nu_{\mu L}-\nu_{\tau L}}{\sqrt{2}}\equiv \nu_{-}$ and
$\frac{\nu_{\mu R}-\nu_{\tau R}}{\sqrt{2}}\equiv \nu_{s}$. The seesaw
mechanism now can
be applied to the remaining $2\times 2$ matrix to yield the following
light neutrino mass matrix in the original $3\times 3$ basis.
\begin{eqnarray}
{\cal M}^{(0)} = \left(\begin{array}{ccc} 0 & m & m\\ m & 0 & 0\\ m & 0& 0
\end{array}\right);
\end{eqnarray}
with $m= \frac{m_0m_{11}}{\sqrt{2}M}$.
This matrix on diagonalization leads to the strict bimaximal MNS matrix
given above. We thus see that in the strict bimaximal limit there are two
massless neutrinos: one, $\nu_{-}$ is an active left-handed neutrino
and the other, $\nu_{s}$ is a
right handed neutrino, that has no weak interactions of Fermi
strength. The latter can therefore play the role of a sterile
neutrino.

The special case of the mass matrix in equation (8) ${\cal M}^{(0)}$)
predicts an inverted spectrum for neutrinos and oscillation of atmospheric
muon neutrinos dictated by the $\Delta m^2_{ATMOS}= 2 m^2$; it is however
not realistic since it leads to $\Delta m^2_{\odot}=0$ and therefore 
predicts no oscillation of the solar neutrinos. In order to make
this model useful, we must add small corrections to it. 
Such corrections to the mass matrix can arise once the
$L_e-L_{\mu}-L_{\tau}$ as well as the permutation symmetries are broken.
 In that case both the sterile neutrino mass as well as
its mixings with the active neutrinos can arise. In the next section, we
give an example of a model for this case.

To study such realistic situations,
let us consider the case, when $L_e-L_{\mu}-L_{\tau}$ and
$S_{2R}$ are broken
softly in the Higgs sector. An example of a mass matrix with this symmetry
breaking can be of the 
form (in the basis $(\nu_e, \nu_{+}, \nu_{-}, \nu_s)$, where
$\nu_{\pm}\equiv \frac{\nu_{\mu}\pm \nu_{\tau}}{\sqrt{2}}$):
 \begin{eqnarray}
{\cal M}^{(0)}+{\cal M}^{(1)}~=~ \left(\begin{array}{cccc} 0 &\sqrt{2} m
& 0 &
0\\
\sqrt{2}m & \delta_3 &  0 &\delta_2\\ 0 & 0 & 0 &0 \\ 0 &
\delta_2 & 0 & \delta_1 \end{array}\right)
\end{eqnarray}
 (Note that it still respects the permutation symmetry
$S_{2L}$), 
Diagonalization of this mass matrix in the approximation $\delta_2\ll m\ll
\delta_1$, 
leads to the following form for the
generalized $4\times 4$ MNS matrix $\tilde{U}$:
\begin{eqnarray}
\tilde{U}~=~\left(\begin{array}{cccc}
c & s & 0 & \epsilon \\ -\frac{s}{\sqrt{2}} & \frac{c}{\sqrt{2}} &
\frac{1}{\sqrt{2}} & \frac{\alpha}{\sqrt{2}}\\ \frac{s}{\sqrt{2}}
& -\frac{c}{\sqrt{2}} & \frac{1}{\sqrt{2}} & -\frac{\alpha}{\sqrt{2}}\\
\alpha s & -\alpha c & 0 & 1 \end{array} \right)
\end{eqnarray}
where $\alpha = \frac{\delta_2}{\delta_1}$ and $\epsilon=\frac{\delta_2
m}{\delta^2_1}$.
This form leads to  $\Delta m^2_{ATMOS}\simeq 2m^2$ as before; however,
now we have $\Delta m^2_{\odot}\simeq \sqrt{2}(2m\delta_3 -\frac{2
m\delta^2_2}{\delta_1} )$
making it possible to understand solar neutrino deficit in terms of
the large mixing angle MSW solution. Clearly a cancellation
between $\delta_3$ and $\delta^2_2/\delta_1$ is required for this
purpose.  To see the extent of fine tuning, note that, we get for the LSND
mixing $\theta_{LSND}\simeq \alpha\epsilon\equiv
\frac{\delta^2_2m}{\sqrt{2}\delta^3_1}$. If we choose $\delta_1 \sim 0.6 $
eV (the LSND mass) and LSND mixing to be $0.02$, then we get
$\delta^2_2/\delta_1\simeq 0.16$; so to get the right order of magnitude
for the $\Delta m^2_{\odot}$, we need fine tuning at the level of one per
cent or so. The extent of fine tuning needed is less severe for lower
values of $\Delta m^2_{LSND}$ e.g. for $\delta_1 \simeq 0.2$ eV, the
required fine tuning is at the level of 10\%. In this model $U_{e3}=0$ but
\begin{eqnarray} U_{e4}\simeq
-\frac{\delta_2 m}{\delta^2_1} \approx 0.04 -0.06,
\end{eqnarray}
which may be accessible to future high precision disappearance searches
for $\nu_e$'s. This may qualify as a realistic 3+1 scenario.

\section{Theoretical origin of almost bimaximal form and sterile neutrino
mass}

In this section, we discuss theoretical schemes where by extending the
standard model to include the right handed neutrinos and several extra
Higgs fields, one can obtain the almost bimaximal mass matrix that leads
to the MNS matrix discussed in the previous section. One
may use other alternative schemes such as the use of higher dimensional
terms that include Higgs fields that break both
the $L_e-L_{\mu}-L_{\tau}$ and $S_{2R}$ symmetry. Here we give an example
of only the first kind.

The radiative scheme uses only the standard model
gauge group with the right handed
neutrinos to implement the seesaw mechanism and to get the massless
sterile neutrino, as has already been discussed. The leptonic
multiplets $(L_{\mu}, L_{\tau})$ are $S_{2L}$ multiplets, singlets
$(\mu_R,\tau_R)$ and
$(\nu_{\mu R}, \nu_{\tau R})$ are the $S_{2R}$ multiplets. The extra Higgs
fields
included are: iso-singlet singly (positive) charged fields $\eta^{+,-}_2$,
$\eta^{+,-}_0$,
iso-singlet doubly (positive) charged fields $H^{+,+}_2$, $H^{+,+}_0$,
where the
superscripts correspond to the $S_{2L}\times S_{2R}$ quantum numbers
($+$ means even and $-$ means odd under permutation) and
subscripts to $L_e-L_{\mu}-L_{\tau}$ quantum numbers. We also include an
extra standard model like doublet $\phi^{+,-}_2$ in addition to the 
usual Higgs doublet $\phi_1$. The standard model
Lagrangian is then augmented by the inclusion of the then following terms:
\begin{eqnarray}
{\cal L'}=~M_1 \eta_0\phi_1\phi_2 + f_1\eta_2\ell_{+R}\nu_{-R}
+f_2\eta_0e_R\nu_{-R}+ h_1H_2(\ell_{+R}\ell_{+R} + a \ell_{-R}\ell_{-R})
\nonumber\\ + H_0e_R\ell_{+R}+ M H^*_2\eta_0\eta_2
+M'H^*_0\eta_0\eta_0+\mu M H^*_2H_0 + h.c.
\end{eqnarray}
where we have denoted $S_{2L}$ even and odd lepton doublet combinations by
$\ell_{\pm}$, $S_{2R}$ odd RH neutrino combination as $\nu_{-R}$
(previously called $\nu_s$) and
$S_{2R}$ even and odd combinations of the charged leptons as $\ell_{\pm
R}$. We assume all the new fields to be in the TeV scale. This is an extra
assumption which is at the same level as , say , the $\mu$ problem of
supersymmetry and presumably can be solved once the nature of physics
beyond the standard model becomes more clear.

 Using these interactions, one gets at the two loop level\cite{babu}
(for a generaic diagram, see Fig. 1) $\nu_{-R}\nu_{-R}$ (or $\nu_s\nu_s$) 
and
$\nu_{-R}\nu_{+L}$ as well as
$\nu_{+L}\nu_{+L}$ contributions to the $4\times 4$ neutrino mass
matrix. The magnitude of these contributions are $\sim
\frac{f^2h \mu}{(16\pi^2)^2}$ where we assume $\mu$ to be of order of the
weak
scale. For $f\simeq h\simeq 10^{-2}$, this gives the new contributions to
the neutrino mass matrix of order .1 - 1 eV, which can then lead to a
realsitic picture for all three neutrino oscillations (solar, atmospheric
and LSND). The charged lepton sector can be made diagonal using
appropriate number of Higgs fields which we do not discuss here.

In conclusion, it is pointed out in this brief note that there may be an
intimate connection between the exact bimaximal neutrino mixing matrix and
the possible existence of a light sterile neutrino within the conventional
seesaw picture for neutrino masses. This connection arises from new
symmetries of the neutrino mass matrix that yield the exact
bimaximal MNS matrix. The resulting scenario is a 3+1
scheme for understanding the present neutrino oscillation data, which, 
though highly constrained at the moment, may still be a viable
possibility. A crucial prediction of these models is the inverted spectrum
for neutrinos. 

This work is supported by the National Science Foundation Grant
No. PHY-0099544. It was presented at one of the P4 sessions of the
Snowmass 2001 workshop; he would like to thank the organizers for a very
stimulating workshop.

\noindent{\bf Note added:} It is possible to show that there is a
connection between the near bimaximal neutrino mixing and the existence of
a light sterile neutrino if the theory has a smaller symmetry
$U(1)_{L_e-L_{\mu}-L_{\tau}}\times S_{2R}$ symmetry. The $S_{2R}$ symmetry
acts only on the right handed neutrinos. As a result the charged lepton
sector is as in the standard model. This is the subject of a forthcoming
article.

\begin{figure}[htb] \begin{center}
\epsfxsize=7.5cm
\epsfysize=7.5cm
\mbox{\hskip -1.0in}\epsfbox{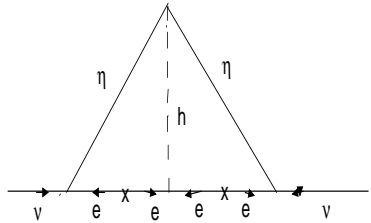}
\caption{Two loop diagram for sterile neutrino mass and mixing
\label{Fig.1}}
\end{center}
\end{figure}

\end{document}